# Transient response of spin Peltier effect revealed by lock-in thermoreflectance measurement

Takumi Yamazaki[*,1], Ryo Iguchi[*,2,†], Tadakatsu Ohkubo[2], Hosei Nagano[1], and Ken-ichi Uchida[2,3,4,5]

[1] *Department of Mechanical Systems Engineering, Nagoya University, Nagoya 464-8601, Japan*

[2] *National Institute for Materials Science, Tsukuba 305-0047, Japan*

[3] *Institute for Materials Research, Tohoku University, Sendai 980-8577, Japan*

[4] *Center for Spintronics Research Network, Tohoku University, Sendai 980-8577, Japan*

[5] *Department of Mechanical Engineering, The University of Tokyo, Tokyo 113-8656, Japan*

[Abstract]

Transient response of the spin Peltier effect (SPE) in a Pt/yttrium iron garnet junction system has been investigated by means of a lock-in thermoreflectance method. We applied an alternating charge current to the Pt layer to drive SPE through the spin Hall effect, and measured the AC response of the resultant SPE-induced temperature modulation at frequencies ranging from 10 Hz to 1 MHz. We found that the SPE-induced temperature modulation decreases with increasing the frequency when the frequency is >1 kHz. This is a characteristic feature of SPE revealed by the high frequency measurements based on the lock-in thermoreflectance, while previous low frequency measurements showed that the SPE signal is independent of the frequency. We attribute the decrease of the temperature modulation to the length scale of the SPE-induced heat current; by comparing the experimental results with one-dimensional heat conduction calculations, the length scale of SPE is estimated to be 0.94 μm.

[*] These authors contributed equally to this work.
[†] IGUCHI.Ryo@nims.go.jp

Thermoelectric technology has gained attention because of interests for energy harvesting and thermal management. Recently, a new route for thermoelectric conversion has been opened by using the spin degree of freedom of electrons [1,2]. The spin-mediated thermoelectric generation phenomena, such as the spin Seebeck effect [3-6] and anomalous Nernst effect [7-9], have been widely studied in the field of spin caloritronics. In addition to the thermoelectric generation, the research for spin-mediated temperature manipulation is becoming active. The representative effects are the spin Peltier effect (SPE) [10-16] and anomalous Ettingshausen effect (AEE) [17], which are the reciprocal of the spin Seebeck and anomalous Nernst effects, respectively.

SPE induces a heat current from a spin current in a magnetic material. In combination with the charge-to-spin current conversion, *i.e.*, the spin Hall effect in a conductor attached to the magnetic material, SPE works as a thermoelectric converter [10]. In a junction comprising the magnetic material and conductor, a charge current $\mathbf{J}_c$ applied to the conductor layer induces a transverse spin current with the spin polarization $\sigma$ [18]. The spin current is injected into the magnetic layer via the interfacial exchange coupling between electron spins and the thermally-activated magnon dynamics, and propagates from the interface with accompanying the heat current $\mathbf{J}_q^{SPE}$ [Fig. 1(a)]. The symmetry of SPE is described by

$$\mathbf{J}_q^{SPE} \propto (\sigma \cdot \mathbf{M})\mathbf{n}, \quad (1)$$

where $\mathbf{M}$ denotes the magnetization in the magnetic layer and $\mathbf{n}$ the normal vector to the junction interface [10,12,19]. When $\mathbf{M} \perp \mathbf{n}$, SPE satisfies $\mathbf{J}_q^{SPE} \propto \mathbf{J}_c \times \mathbf{M}$, which is the same symmetry as AEE, appearing in ferromagnetic conductors [Fig. 1(b)] [17]. The transverse thermoelectric conversion based on SPE allows us to heat or cool a large plane with a simple form. SPE has been studied in some different material systems by means of thermocouples [10,13], heat-flux sensors [15], and lock-in thermography (LIT) [11,12,14,16], which clarify the spin/charge current dependence, magnetic field dependence, and spatial distribution of the SPE-induced temperature modulation.

However, the response time of SPE, the time constant of the SPE-induced temperature modulation to reach the steady state, is not experimentally verified yet. The response time is important not only for characterizing and designing SPE-based devices but also for investigating the physical origin of SPE because it is determined by the characteristic length scale of a heat current induced by SPE, which is still under debate [16,20,21]. This is in sharp contrast to the AEE case because the AEE-induced heat current $\mathbf{J}_q^{AEE}$ uniformly exists in ferromagnets and its length scale is determined by the system dimension [17]. To reveal the transient behavior of SPE, a fast measurement within a timescale of thermal propagation is necessary. Although the conventional approach based on LIT is powerful for measuring SPE, this method cannot be used for ultrafast measurements because of the limitation on framerate of an infrared camera. The direct contact method using thermometers is disturbed by thermal resistance between thermometer and sample and thermal capacitance of thermometers [13]. Therefore, to determine the response time of SPE, a different approach is necessary.

In this study, we have investigated the transient response of SPE in a Pt/yttrium iron garnet (YIG) junction system by using a thermoreflectance method, where the temperature modulation is measured through the change in the reflectivity of the sample surface [22]. In a similar manner to the previous studies [11,14,23], we performed the thermal measurements in the frequency domain; reflectivity oscillation synchronized with alternating thermal excitation was measured with changing the excitation frequency. This is often called the lock-in thermoreflectance (LITR) method [24]. In the following, we show the validity of the LITR-based SPE detection by systematically performing the charge current and magnetic field dependent measurements. Then, we reveal the transient response of SPE by changing the frequency from 10 Hz to 1 MHz (note that the maximum frequency of the conventional LIT method is ~100 Hz [14]). Finally, we discuss the length scale of SPE by comparing the obtained frequency responses with model calculations.

The sample configuration is as follows. A 500-μm-wide and 9.0-nm-thick Pt strip was formed on a single-crystalline YIG (111) layer with a thickness of 114 μm. The YIG layer with the exact composition of $Bi_{0.04}Y_{2.96}Fe_5O_{12}$ was grown on a single-crystalline gadolinium gallium garnet (111) substrate by liquid phase epitaxy, which is the same as the YIG samples used for the LIT-based SPE measurements [11,12,16]. To measure the temperature on the Pt/YIG interface using the LITR method, the sample is covered by a 50.6-nm-thick Au transducer layer showing a large thermoreflectance coefficient, *i.e.*, temperature derivative of reflectivity. For electrical insulation between the Pt and Au layers, $SiO_2$ and $Al_2O_3$ layers were formed on the Pt layer before forming the Au layer, where the insulation is confirmed by the fact that the four-probe resistance of the Pt strip not changed before and after depositing the Au layer. The layer structure of our sample is shown in Fig. 1(e) and the detailed preparation procedures are summarized in Ref. 25.

Figure 1(c) shows a schematic diagram of the LITR measurement. The reflectivity change due to SPE-induced temperature modulation is measured by irradiating the sample with a focused continuous-wave green (532 nm) laser light and monitoring the reflected light intensity, where the spot radius is ~10 μm and the power at the sample surface is estimated to be 3.8 mW. To excite SPE, a sinusoidal AC charge current with the frequency $f$ is applied to the Pt layer. The charge current is converted into a spin current via the spin Hall effect, where the $\sigma$ direction of the spin current is alternately reversed. This AC spin current induces periodic temperature oscillation with the same frequency near the Pt/YIG interface due to SPE [see Eq. (1) and Ref. 11], which eventually modulates the reflectivity of the transducer layer through thermal conduction. The resultant AC component of the reflected light intensity is measured by using photodetector and lock-in amplifier [25]; we obtain the amplitude $I_{AC}$ and phase $\phi$ for the reflected light intensity change for the first harmonic response. Then, $I_{AC}$ is converted into the amplitude of the temperature modulation $A$ through $A = (1/R)I_{AC}/I_{DC}$, where $R$ ($= 2.82 \times 10^{-4}$ K$^{-1}$ [26]) denotes the thermoreflectance coefficient and $I_{DC}$ DC intensity of the incident light. Importantly, we can measure the temperature modulation due to thermoelectric effects separated from Joule heating because Joule heating ($\propto \mathbf{J}_c^2$) induces only the second harmonic component (with its amplitude $A_{2f}$), which can also be detected through the lock-in amplifier [Fig. 1(d)]. In this study, we concentrate on the two components: the $H$-odd component of the first harmonic component, $A_{odd} = |A(+H)e^{i\phi(+H)} - A(-H)e^{i\phi(-H)}|/2$, and the Joule-heating-induced component, $A_{Joule} = A_{2f}(+H)$, where $H$ denotes the magnitude of the magnetic field $\mathbf{H}$, applied in-plane to the YIG film across the Pt strip [27]. $A_{odd}$ contains the SPE contribution because the SPE-induced temperature modulation changes its sign with respect to the reversal of the magnetization of the YIG layer and $\mathbf{m}$ follows $\mathbf{H}$ because of negligibly small hysteresis in this experiment. We note that the laser light is ellipsoidally polarized but no polarization dependence is observed as our setup comprises non-polarizing optical components only. The measurements were performed at room temperature and atmospheric pressure.

First, we confirmed the charge current and magnetic field dependences of the LITR signals. Figure 2(a) shows the $J_c^{AC}$ dependence of $A_{odd}$ and $A_{Joule}$ at $\mu_0|H| = 30$ mT at various frequencies, where $J_c^{AC}$ denotes the amplitude of the AC charge current applied to the Pt layer. $A_{odd}$ ($A_{Joule}$) increases linearly (quadratically) with $J_c^{AC}$, confirming that its origin is the thermoelectric effect (Joule heating). Figure 2(b) shows the $\mu_0|H|$ dependence of $A_{odd}$. We found that the $A_{odd}$ value increases with increasing $H$ and its magnitude is saturated for $\mu_0|H| > 10$ mT. This behavior shows a good agreement with the magnetization curve of YIG. The temperature-modulation amplitude over the charge current density observed here is estimated to be $A_{odd}/j_c = 3.6 \times 10^{-13}$ Km$^2$A$^{-1}$, which is comparable to the previously-reported value obtained by the LIT measurements: $A_{odd}/j_c = 3.3 \times 10^{-13}$ Km$^2$A$^{-1}$ [11] (note that a correction factor of $\pi/4$ is applied to the previous LIT result because the sinusoidal wave amplitude of the temperature change is divided by the rectangular wave amplitude of the charge current in the LIT-based studies). The above results confirm that the LITR signals are attributed to the SPE-induced temperature modulation.

Figure 3(a) shows the $f$ dependence of $A_{odd}$ for the Pt/YIG system (blue circles). At low frequencies ($f$ < 1 kHz), the amplitude of the SPE signal is almost constant, as revealed by the LIT measurements [14]. Interestingly, at high frequencies ($f$ > 1 kHz), the signal amplitude gradually decreases with increasing $f$, showing the transient response due to SPE. Considering that the response time $\tau$ is obtained through $f_c = (2\pi\tau)^{-1}$, where $f_c$ is defined by the frequency at which the signal amplitude decreases to -3dB of the maximum value, $\tau$ is estimated to be 2.5 μs for SPE in our Pt/YIG system. Note that the $f$ dependence of $J_c^{AC}$ is irrelevant to the temperature decrease as it is found to be constant in this frequency range as shown in the inset of Fig. 3(a), where the values are estimated from the AC voltage difference $V_{sample}$ inside the strip [see Fig. 1(c)] at each frequency and the low-frequency resistance determined by four-probe measurements.

The transient response of the SPE signal is different from that of other effects. In Fig. 3(a), we also show the $f$ dependence of the temperature modulation due to Joule heating (red squares). The decrease for the Joule heating signal is much larger and faster than that for SPE, which can be attributed to the difference in the heat source distribution [11]. Importantly, we found that the transient response of the SPE signal is different also from that of the AEE signal, although SPE and AEE exhibit the similar thermoelectric conversion symmetry for in-plane magnetized conditions [17]. To demonstrate the difference, we measured the $f$ dependence of $A_{odd}$ for AEE (gray diamonds) by using the sample in which the Pt/YIG layer is replaced with a 13.3-nm-thick Ni film, where the AEE signal appears in the Ni layer instead of the SPE signal. As shown in Fig. 3(a), $A_{odd}$ for AEE maintains its magnitude over the measurement frequency range, indicating much shorter $\tau$ than that of SPE for the Pt/YIG system.

The difference of the frequency response between SPE and AEE can be attributed to the difference in the length scale of $\mathbf{J}_q^{SPE}$ and $\mathbf{J}_q^{AEE}$ when the decoupling between magnon and phonon temperatures can be neglected for SPE. This assumption is justified at room temperature; the magnon-phonon relaxation time is expected to be in the order of picoseconds [20,28], and thus does not affect our experimental results, where the maximum time scale of our experiment is >1 μs. Let us compare the length scale and corresponding $\tau$ value. When a uniform heat current propagates in the distance of $l$ in a one-dimensional system, $\tau$ is given by $\rho Cl^2/(2\kappa)$, where $C$, $\kappa$, and $\rho$ are the heat capacity, thermal conductivity, and density, respectively. Therefore, larger $l$ results in larger $\tau$. Note that when the timescale corresponding to the lock-in frequency is shorter than $\tau$, the LITR signal decreases by reflecting the transient response. In the case of AEE, the uniform $\mathbf{J}_q^{AEE}$ exists over the thickness direction and, by using thermal properties for Ni in Table I, $\tau = 3.8$ ps ($f_c = 42$ GHz) is obtained. Since the AEE signal can reach the steady state in a very short time scale, $A_{odd}$ for AEE in our Ni film shows the constant magnitude for $f$ < 1 MHz. In contrast, if the SPE-induced temperature modulation requires a substantial time to reach the steady state due to the long length scale, the reduction of the SPE signal may appear in the high $f$ range. This means that, by analyzing the $f$ dependence of the SPE signal, we can determine the length scale of the SPE-induced heat current.

To quantitatively interpret the experimental data, we analyze the transient response of SPE with an assumption that the magnitude of $\mathbf{J}_q^{SPE}$ decays exponentially from the Pt/YIG interface with the characteristic length $l_q^{SPE}$. We note that the heat current generation can be regarded as one-dimensional transport because the Pt strip width is much larger than $l_q^{SPE}$ [25]. We solved the one-dimensional heat equation in the multilayer structure [Fig. 3(c)] and derived the frequency response [25]. The interfacial thermal conductance is set to $2.79 \times 10^8$ Wm$^{-2}$K$^{-1}$ for the interface of Pt/YIG [28] and $1 \times 10^8$ Wm$^{-2}$K$^{-1}$ for the other interfaces [31,32].

Figure 3(b) shows that the calculated $f$ dependence of $A_{odd}$ on the Au surface induced by SPE for various $l_q^{SPE}$ values. It is clear that larger $l_q^{SPE}$ results in lower $f_c$. By fitting the experimental result with $l_q^{SPE}$ being the adjustable parameter, we obtained $l_q^{SPE} = 0.94$ μm. This value is similar to but smaller than that obtained from the measurements of the spin Seebeck effect and non-local magnon transport [13,33-46], which may be due to the different length scale of spin-current excited magnons in SPE from that of thermally-excited magnons in the spin Seebeck effect. The exact comparison of $l_q^{SPE}$ to the length scale of the spin Seebeck effect is a remaining task for understanding the reciprocal relationship in the spin-heat current conversion. The μm-scale $l_q^{SPE}$ indicates that, in Pt/YIG systems, the bulk magnon conduction plays a major role for determining the magnitude of the temperature modulation induced by SPE.

In conclusion, we demonstrated the LITR-based SPE detection and revealed the transient response of the temperature modulation due to SPE. The AC response of the SPE signal in the Pt/YIG system was found to decrease with increasing $f$, showing the response time of 2.5 μs. The obtained $f$ dependence of the SPE-induced temperature modulation is well reproduced by the model calculation based on the one-dimensional heat equation. By assuming the exponential decay of magnons, we show that the heat current induced by SPE exists over 0.94 μm in the YIG layer from the Pt/YIG interface. We also found that, despite the similar thermoelectric conversion symmetry, the transient response of SPE is different from that of AEE in an in-plane magnetized thin film, where the $f$ dependence of the AEE signal shows no decrease according to the ultrafast response time determined by the sample dimension. As demonstrated in this study, the LITR method can be a useful tool for investigating the physics of thermoelectric and thermo-spin conversion phenomena and for evaluating the transient response of spin-caloritronic devices.

TABLE I. Materials parameters used for calculation. $C$, $\kappa$, and $\rho$ denote the specific heat, thermal conductivity, and density, respectively.

|      | $C$ (J/kg/K) | $\kappa$ (W/m/K) | $\rho$ (kg/m$^3$) |
|------|--------------|------------------|-------------------|
| Au   | 129[a]       | 317[a]           | 19300[a]          |
| Al$_2$O$_3$ | 779[a] | 35[a]            | 3970[a]           |
| SiO$_2$ | 745[a]    | 1.4[a]           | 2196[a]           |
| Pt   | 133[a]       | 72[a]            | 21450[a]          |
| YIG  | 570[b]       | 7.4[c]           | 5170[b]           |
| Ni   | 440[a]       | 90.7[a]          | 8900[a]           |

[a] Ref. [29].
[b] Ref. [28].
[c] Ref. [30].

The authors thank T. Yagi, Y. Yamashita, N. Ogawa, T. Ishizaki, R. Hisatomi, and G.E.W. Bauer for valuable discussions and J. Uzuhashi, H. Nakayama, and M. Isomura for technical supports. This work was supported by CREST "Creation of Innovative Core Technologies for Nano-enabled Thermal Management" (JPMJCR17I1) from JST, Japan, Grant-in-Aid for Scientific Research (B) (JP19H02585), Grant-in-Aid for Early-Career Scientists (JP18K14116), and Grant-in-Aid for Scientific Research (S) (JP18H05246) from JSPS KAKENHI, Japan, and Joint Research Hub Program from NIMS, Japan. T.Y. is supported by Grant-in-Aid for JSPS Fellows (JP18J23465).

**REFERENCES**
[1] G. E. W. Bauer, E. Saitoh, and B. J. van Wees, Nat. Mater. **11**, 391 (2012).
[2] S. R. Boona, R. C. Myers, and J. P. Heremans, Energy Environ. Sci. **7**, 885 (2014).
[3] K. Uchida, S. Takahashi, K. Harii, J. Ieda, W. Koshibae, K. Ando, S. Maekawa, and E. Saitoh, Nature **455**, 778 (2008).
[4] K. Uchida, J. Xiao, H. Adachi, J. Ohe, S. Takahashi, J. Ieda, T. Ota, Y. Kajiwara, H. Umezawa, H. Kawai, G. E. W. Bauer, S. Maekawa, and E. Saitoh, Nat. Mater. **9**, 894 (2010).
[5] C. M. Jaworski, J. Yang, S. Mack, D. D. Awschalom, J. P. Heremans, and R. C. Myers, Nat. Mater. **9**, 898 (2010).
[6] K. Uchida, H. Adachi, T. Ota, H. Nakayama, S. Maekawa, and E. Saitoh, Appl. Phys. Lett. **97**, 172505 (2010).
[7] T. Miyasato, N. Abe, T. Fujii, A. Asamitsu, S. Onoda, Y. Onose, N. Nagaosa, and Y. Tokura, Phys. Rev. Lett. **99**, 086602 (2007).
[8] Y. Sakuraba, K. Hasegawa, M. Mizuguchi, T. Kubota, S. Mizukami, T. Miyazaki, and K. Takanashi, Appl. Phys. Express **6**, 033003 (2013).
[9] A. Sakai, Y. P. Mizuta, A. A. Nugroho, R. Sihombing, T. Koretsune, M. Suzuki, N. Takemori, R. Ishii, D. Nishio-Hamane, R. Arita, P. Goswami, and S. Nakatsuji, Nat. Phys. **14**, 1119 (2018).

[10] J. Flipse, F. K. Dejene, D. Wagenaar, G. E. W. Bauer, J. B. Youssef, and B. J. van Wees, Phys. Rev. Lett. **113**, 027601 (2014).

[11] S. Daimon, R. Iguchi, T. Hioki, E. Saitoh, and K. Uchida, Nat. Commun. **7**, 13754 (2016).

[12] S. Daimon, K. Uchida, R. Iguchi, T. Hioki, and E. Saitoh, Phys. Rev. B **96**, 024424 (2017).

[13] R. Itoh, R. Iguchi, S. Daimon, K. Oyanagi, K. Uchida, and E. Saitoh, Phys. Rev. B **96**, 184422 (2017).

[14] R. Iguchi and K. Uchida, Jpn. J. Appl. Phys. **57**, 0902B6 (2018).

[15] A. Sola, V. Basso, M. Kuepferling, C. Dubs, and M. Pasquale, Sci. Rep. **9**, 2047 (2019).

[16] S. Daimon, K. Uchida, N. Ujiie, Y. Hattori, R. Tsuboi, and E. Saitoh, arXiv:1906.01560 (2019).

[17] T. Seki, R. Iguchi, K. Takanashi, and K. Uchida, Appl. Phys. Lett. **112**, 152403 (2018).

[18] J. Sinova, S. O. Valenzuela, J. Wunderlich, C. H. Back, and T. Jungwirth, Rev. Mod. Phys. **87**, 1213 (2015).

[19] Y. Ohnuma, M. Matsuo, and S. Maekawa, Phys. Rev. B **96**, 134412 (2017).

[20] L. J. Cornelissen, K. J. H. Peters, G. E. W. Bauer, R. A. Duine, and B. J. van Wees, Phys. Rev. B **94**, 014412 (2016).

[21] V. Basso, E. Ferraro, A. Magni, A. Sola, M. Kuepferling, and M. Pasquale, Phys. Rev. B **93**, 184421 (2016).

[22] W. Claeys, S. Dilhaire, V. Quintard, J. P. Dom, and Y. Danto, Qual. Reliab. Eng. Int. **9**, 303 (1993).

[23] N. Roschewsky, M. Schreier, A. Kamra, F. Schade, K. Ganzhorn, S. Meyer, H. Huebl, S. Geprägs, R. Gross, and S. T. B. Goennenwein, Appl. Phys. Lett. **104**, 202410 (2014).

[24] C. A. Paddock and G. L. Eesley, J. Appl. Phys. **60**, 285 (1986).

[25] See Supplemental Material at [URL will be inserted by publisher] for sample fabrication, noise evaluation, position dependence of the SPE signals, and calculations of the $f$ dependence of the SPE signals.

[26] R. B. Wilson, B. A. Apgar, L. W. Martin, and D. G. Cahill, Opt. Express **20**, 28829 (2012).

[27] The H-even component of the first harmonic signal originates from the Peltier effect, electromagnetic induction, and Joule heating contamination due to small DC offset beyond the control precision of the current source.

[28] M Schreier, A. Kamra, M. Weiler, J. Xiao, G. E. W. Bauer, R. Gross, and S. T. B. Goennenwein, Phys. Rev. B **88**, 094410 (2013).

[29] D. R. Lide, CRC Handbook of Chemistry and Physics, 85th ed. (CRC Press, 2005).

[30] G. A. Slack and D. W. Oliver, Phys. Rev. B **4**, 592 (1971).

[31] J. Zhu, D. Tang, W. Wang J. Liu, K. W. Holub, and R. Yang, J. Appl. Phys. **108**, 094315 (2010).

[32] P. E. Hopkins, J. R. Serrano, L. M. Phinney, S. P. Kearney, T. W. Grasser, and C. T. Harris, J. Heat Transfer **132**, 081302 (2010).

[33] M. Agrawal, V. I. Vasyuchka, A. A. Serga, A. Kirihara, P. Pirro, T. Langner, M. B. Jungfleisch, A.V. Chumak, E. T. Papaioannou, and B. Hillebrands, Phys. Rev. B **89**, 224414 (2014).

[34] S. M. Rezende, R. L. Rodríguez-Suárez, R. O. Cunha, A. R. Rodrigues, F. L. A. Machado, G. A. F. Guerra, J. C. L. Ortiz, and A. Azevedo, Phys. Rev. B **89**, 014416 (2014).

[35] A. Kehlberger, U. Ritzmann, D. Hinzke, E. Guo, J. Cramer, G. Jakob, M. C. Onbasli, D. H. Kim, C. A. Ross, M. B. Jungfleisch, B. Hillebrands, U. Nowak, and M. Kläui, Phys. Rev. Lett. **115**, 096602, (2015).

[36] H. Jin, S. R. Boona, Z. Yang, R. C. Myers, and J. P. Heremans, Phys. Rev. B **92**, 054436 (2015).

[37] T. Kikkawa, K. Uchida, S. Daimon, Z. Qiu, Y. Shiomi, and E. Saitoh, Phys. Rev. **B** 92, 064413 (2015).

[38] L. J. Cornelissen, J. Liu, R. A. Duine, J. B. Youssef, and B. J. van Wees, Nat. Phys. **11**, 1022 (2015).

[39] B. L. Giles, Z. Yang, J. S. Jamison, and R. C. Myers, Phys. Rev. B **92**, 224415 (2015).

[40] S. T. B. Goennenwein, R. Schlitz, M. Pernpeintner, K. Ganzhorn, M. Althammer, R. Gross, and H. Huebl, Appl. Phys. Lett. **107**, 172405 (2015).

[41] E.-J. Guo, J. Cramer, A. Kehlberger, C.A. Ferguson, D.A. MacLaren, G. Jakob, and M. Kläui, Phys. Rev. X **6**, 031012 (2016).

[42] T. Hioki, R. Iguchi, Z. Qiu, D. Hou, K. Uchida, and E. Saitoh, Appl. Phys. Express **10**, 073002 (2017).

[43] X.J. Zhou, G.Y. Shi, J.H. Han, Q.H. Yang, Y.H. Rao, H.W. Zhang, L.L. Lang, S.M. Zhou, F. Pan, and C. Song, Appl. Phys. Lett. **110**, 062407 (2017).

[44] B. L. Giles, Z. Yang, J. S. Jamison, J. M. Gomez-Perez, S. Vélez, L. E. Hueso, F. Casanova, and R. C. Myers, Phys. Rev. B **96**, 180412(R) (2017).

[45] A. Prakash, B. Flebus, J. Brangham, F. Yang, Y. Tserkovnyak, and J.P. Heremans, Phys. Rev. B **97**, 020408 (2018).

[46] J. S. Jamison, Z. Yang, B. L. Giles, J. T. Brangham, G. Wu, P. C. Hammel, F. Yang, and R. C. Myers, Phys. Rev. B **100**, 134402 (2019).

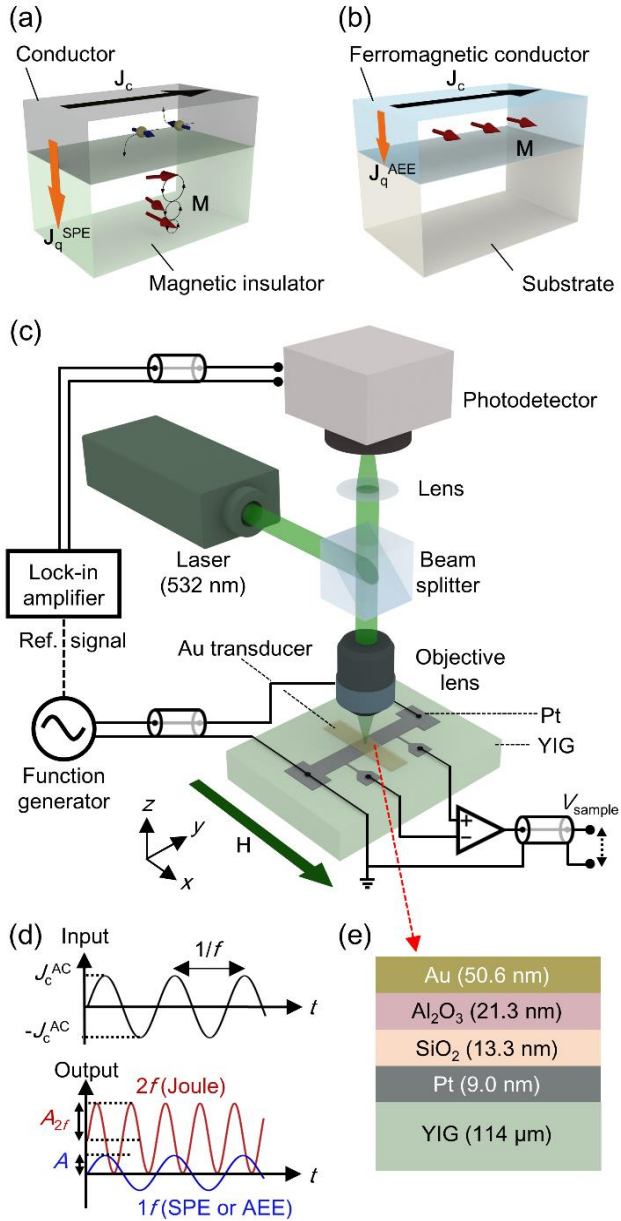

FIG. 1 (a),(b) Schematics of the spin Peltier effect (SPE) in a conductor/magnetic insulator junction system and the anomalous Ettingshausen effect (AEE) in a ferromagnetic metal. **M**, **J**$_c$, and **J**$_q^{SPE(AEE)}$ denote the magnetization vector charge current, and heat current induced by SPE(AEE), respectively. (c) A schematic of the lock-in thermoreflectance (LITR) system used in this study, where **H** denotes the magnetic field. The Pt strip has four electrodes; the outer two are for applying the current and the inner two are for measuring the AC voltage difference $V_{sample}$. Except the connection to the sample electrodes, 50-ohm coaxial cables are used for impedance matching. (d) Expected temporal response of output signals (temperature modulation and intensity of the reflected light) due to the AC charge current. The SPE- and/or AEE-induced signals oscillate at the same frequency as the sinusoidal input whereas the Joule-heating-induced signal oscillates at the second harmonic frequency. (e) Layer structure of the sample used for the SPE measurements. The thickness of each layer was measured by the scanning transmission electron microscopy [25].

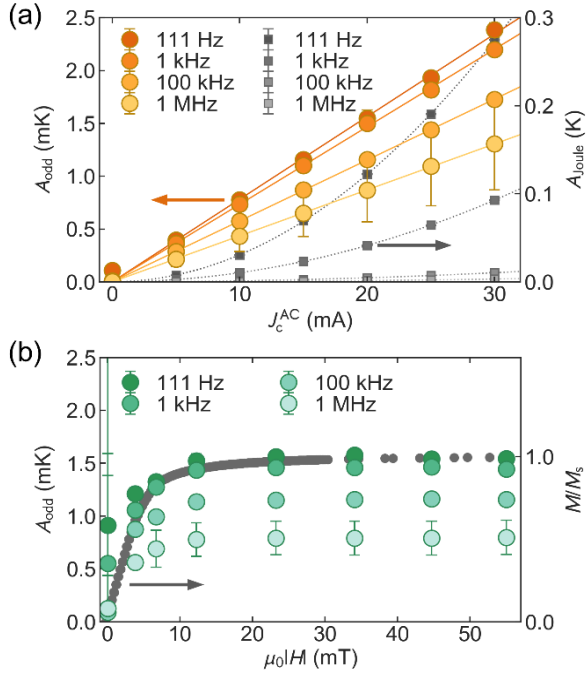

FIG. 2 (a),(b) Charge current ($J_c^{AC}$) and magnetic field ($H$) dependence of the temperature modulation amplitude with the $H$-odd dependence ($A_{odd}$) in the Pt/YIG system for various values of $f$. The $J_c^{AC}$ dependence was measured at $\mu_0|H|$=30 mT and the $\mu_0|H|$ dependence was measured at $J_c^{AC}$=20 mA. In Fig. 2(a), charge current dependence of the temperature modulation amplitude with the Joule-heating-induced component $A_{Joule}$ is also shown. In Fig. 2(b), the magnetization $M$ (normalized by the saturation magnetization $M_s$) curve of YIG measured using a vibrating sample magnetometer is shown for comparison, which shows negligibly small hysteresis. The $J_c^{AC}$ value is estimated from $V_{sample}$ [see Fig. 1(c)] and the sample resistance.

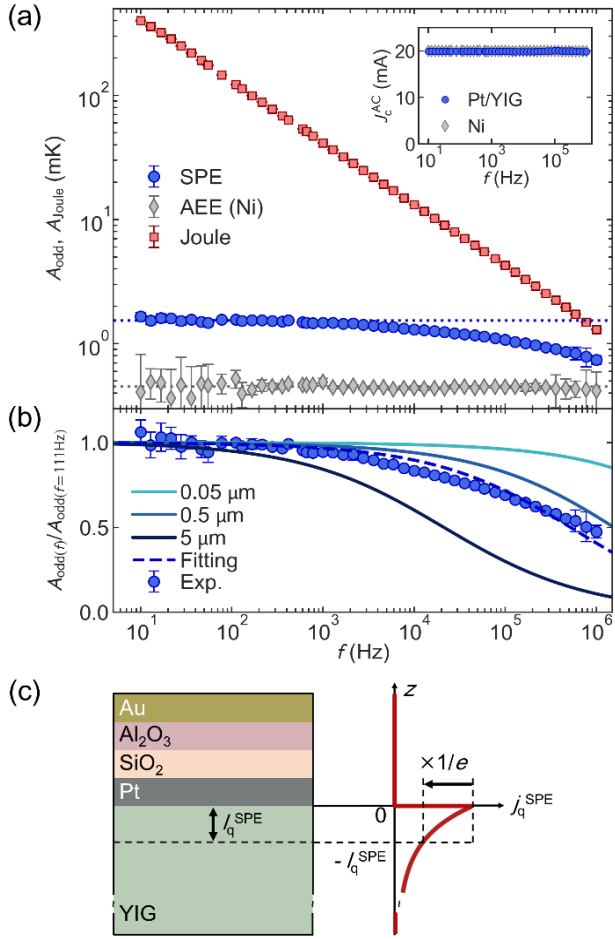

FIG. 3 (a) $f$ dependence of $A_{\text{odd}}$ induced by SPE in the Pt/YIG system (blue circles) and AEE in the Ni system (gray diamonds) and $A_{\text{Joule}}$ induced by Joule heating in the Pt/YIG system (red squares). Note that the dashed lines represent the eye guides, locating at the average values of $A_{\text{odd}}$ in the frequency range from 10 Hz to 1 kHz. The inset shows the $f$ dependence of $J_c^{\text{AC}}$ calculated from $V_{\text{sample}}$ and sample resistance. Note that $J_c^{\text{AC}}$ values for Pt/YIG and Ni samples are almost flat and overlapped in the frequency range. (b) Comparison between the $f$ dependence of SPE calculated with various values of the heat source thickness $l_q^{\text{SPE}}$ in the YIG layer (solid lines and dashed line) and that of the experiment (blue circles). (c) The multilayer model system and the distribution of the SPE-induced heat current density $j_q^{\text{SPE}}$ used for the calculation.